\documentclass[conference]{IEEEtran}

\usepackage{amssymb}
\usepackage{amsmath}
\usepackage{graphicx}

\ifCLASSINFOpdf
\else
\fi

\usepackage{mdwmath}

\usepackage{eqparbox}

\usepackage{fixltx2e}

\usepackage{stfloats}
\begin{document}
%
\title{\huge Analytical Studies of Fragmented-Spectrum Multi-Level OFDM-CDMA Technique in Cognitive Radio Networks}

\author{\IEEEauthorblockN{Farhad Akhoundi, Saeed Sharifi-Malvajerdi, Omid Poursaeed, Jawad A. Salehi \noindent \footnote{The authors are with the Electrical Engineering Department of Sharif University of Technology, Tehran, Iran.\\
Corresponding author: J. A. Salehi, Professor of Electrical Engineering, Sharif University of Technology (e-mail: jasalehi@sharif.edu).}, \textit{Fellow, IEEE}}
\IEEEauthorblockA{School of Electrical Engineering\\
Sharif University of Technology\\
Tehran, Iran \\
Email: akhoundi@alum.sharif.edu,  saeedsharifi, omid{\_}poursaeed and jasalehi@sharif.edu}}


%


\maketitle

\begin{abstract}
\textbf{In this paper, we present a multi-user resource allocation framework using fragmented-spectrum synchronous OFDM-CDMA modulation over a frequency-selective fading channel. In particular, given pre-existing communications in the spectrum where the system is operating, a channel sensing and estimation method is used to obtain information of subcarrier availability. Given this information, some real-valued multi-level orthogonal codes, which are orthogonal codes with values of $\{\pm1,\pm2,\pm3,\pm4, ... \}$, are provided for emerging new users, i.e., cognitive radio users. Additionally, we have obtained a closed form expression for bit error rate of cognitive radio receivers in terms of detection probability of primary users, CR users' sensing time and CR users' signal to noise ratio. Moreover, simulation results obtained in this paper indicate the precision with which the analytical results have been obtained in modeling the aforementioned system.}
\end{abstract}


%
\IEEEpeerreviewmaketitle

\section{Introduction}
\PARstart \noindent {\Huge M}\normalsize
\noindent ODERN wireless communication services require high bandwidth to support multimedia formats. Direct-sequence code-division multiple-access (DS-CDMA) has emerged as a core wireless technology in the third generation and emerging standards. However, conventional CDMA systems are fundamentally limited in their ability to deliver high data rates due to practical issues associated with higher chip rates and complexity issues related to higher inter symbol interference (ISI) \cite{hara1997overview}. Hence, multicarrier (MC) modulation, often called orthogonal frequency-division multiplexing (OFDM), has attracted considerable attention due to its ability to support high rates as well as improving ISI and fading robustness. In OFDM, the high rate stream is divided into a number of parallel lower rate ones that are transmitted over narrowband orthogonal subcarriers. An interesting point of OFDM is that modulation and demodulation can be carried out by Discrete Fourier Transform (DFT) operation.

A hybrid of CDMA and OFDM, namely OFDM-CDMA, is proposed to provide high data rates in typical CDMA systems \cite{chen2009performance}. In OFDM-CDMA, the signature code is applied across a number of orthogonal subcarriers in the frequency domain. The lower data rate provided by each subcarrier is shown clearly in longer symbol and chip interval. System can be properly designed so that each subcarrier encounters flat fading, thereby eliminating channel equalization requirement. Moreover, the system takes the advantage of frequency selectivity for diversity via the different subcarriers.

The main challenge for the deployment of next generation systems in wireless multi-user communications is how to effectively exploit the system resources, such as frequency spectrum and power. Cognitive radio, a technique that allows users to dynamically sense the frequency spectrum, finds the available spectrum bands in a target spectral range, and then transmits without introducing interference to the existing users in this spectral range. Thus, it provides a technique that effectively enhances the usage of system resources \cite{haykin2005cognitive}. In the following subsections, related studies in this area and our contributions are discussed.
\subsection{Related works}
Various schemes have been proposed to facilitate the dynamic spectrum access in such a cognitive radio based network. In \cite{etkin2007spectrum}, a spectrum-sharing problem in an unlicensed band is studied under game-theoretic approaches. In \cite{fan2011joint}, the problem of joint multichannel cooperative sensing and resource allocation is investigated where an optimization problem is formulated, which maximizes the weighted sum of secondary users' throughputs while guaranteeing a certain level of protection for the activities of primary users. In \cite{qu2008cognitive}, a military cognitive radio based multi-user resource allocation framework for mobile ad hoc networks using multi-carrier DS CDMA modulation has been proposed. In this scheme, in particular, the entire spectral range is first sensed, and then the un-used subcarrier bands are assigned to each cognitive radio (CR) user who employs Direct Sequence Spread Spectrum (DS-SS) modulation.

In \cite{wu2008performance}, an interesting dynamic spectrum access method based on non-contiguous carrier interferometry MC-CDMA has been introduced and its performance is evaluated. Particularly, CR users need to deactivate some of their subcarriers to avoid interference to primary users; and orthogonal Carrier Interferometry codes are used to eliminate the loss of orthogonality among spreading codes caused by deactivating subcarriers. However, compared with simple binary Hadamard-Walsh signature codes with values $\{\pm1\}$, orthogonal Carrier Interferometry codes, with length $N$, take complex values $\{e^{\frac{-j2k\pi}{N}n}\}$ for $n^{th}$ chip of the $k^{th}$ signature code that result in complexity of the system, from implementation point of view; also, this scheme does not consider the probabilities of false alarm and detection  and \cite{shoreh2014design} of primary users in computing error probability. In a similar study, the authors in \cite{akhoondi2014resource}, \cite{akhoondi2012multiple}, introduced a partially matched filter to eliminate narrow band interference (NBI), which can be interpreted as interference from primary users. However, detection of NBI was assumed to be ideal, which is far from reality.
\subsection{Contributions}
In this paper, we take advantage of hybrid synchronous OFDM-CDMA modulation and utilize it as CR users' modulation in a cognitive radio-based network to provide high data rate by efficiently exploiting available spectrum bands in a target spectral range while simultaneously offering multiple-access capability.
In comparison to the similar study in \cite{wu2008performance}, we analyze the error probability of the fragmented-spectrum synchronous OFDM-CDMA in terms of detection probability of primary users, CR users' sensing time and CR users' signal to noise ratio. Furthermore, in order to eliminate the loss of orthogonality among spreading codes caused by deactivating subcarriers, we apply real-valued multi-level orthogonal codes, i.e., codes with values of $\{\pm1,\pm2,\pm3,\pm4, ... \}$ instead of complex orthogonal Carrier Interferometry codes. Moreover, We have obtained a closed form expression for bit error rate of the CR users' receiver over frequency-selective fading channel.

In particular, a cognitive radio network with multiple potential subcarriers and multiple CR users is considered. Each CR user carries out its local wideband spectrum sensing with details in \cite{quan2009optimal} to get a test statistic for each subcarrier, and makes a binary decision as to whether primary users are present or not. All CR users forward their decisions to a coordinator which is defined as common receiver in \cite{letaief2009cooperative}. Having fused all the decisions from CR users, the coordinator makes the final estimation as to whether primary users have occupied a subcarrier or not. When a subcarrier is estimated to be busy, the coordinator will deactivate it, and broadcast the modified signature sequences for CR users via a common control channel. Thus, CR users spread their power on free subcarriers and transmit without interfering with primary users.

As discussed in \cite{wu2008performance}, this scheme not only improves performance of the system, but also evades having to design various and complex spectrum sharing methods. In other words, fragmented-spectrum OFDM-CDMA modulation will shift the design criterion to an algorithmically oriented code modification which can be easily implemented through software programming. Also, despite the state-of-the-art cognitive radio systems, which assign each free subcarrier to one CR user, resource allocation using synchronous OFDM-CDMA modulation exploits frequency diversity by spreading power of each CR user on all free subcarriers.

The rest of this paper is organized as follows. In section II, we briefly describe spectrum sensing method used in our system and explain resource allocation scheme based on fragmented-spectrum synchronous OFDM-CDMA. The signal model for CR users is also introduced in this section. Section III presents the performance evaluation of a typical CR receiver. In section IV, we present numerical results and discuss on both analytical and simulation-based outcomes. Section V concludes the paper.
\section{SYSTEM MODEL}
A cognitive radio network is considered with $N$ subcarriers and $K$ secondary communication pairs, with the $k^{th}$ pair including secondary transmitter $k$ and secondary receiver $k$. The subcarriers have a bandwidth of $\Omega $ and are licensed to primary users. The system is assumed to be synchronized, and time is separated into slots, each with a length of $T$. In each slot, the primary user (PU) in a channel is assumed to be either active or idle for the whole time interval $T$. We also assume that channels among primary and secondary users remain unchanged within each slot. In other words, the channel coherence time is assumed to be greater than $T$.

In each time slot, firstly each CR user senses the channel, and makes a binary decision on whether the PU is present or not; and the results will be sent to the coordinator \cite{letaief2009cooperative}. Then, the coordinator modifies and broadcasts cognitive radio users' spreading codes based on the collected information from CR users. Finally, the CR pairs' communication takes place by utilizing modified signature sequences in the rest of the slot. The information exchange between CR users and the coordinator is on a common control channel. Since the code modification and broadcasting phase adds a fixed time overhead, we assume its duration to be zero without loss of generality.

\subsection{ Wideband Spectrum Sensing}
\noindent The task of spectrum sensing is to sense subcarriers and determine spectral holes for opportunistic use. For simplicity, all cognitive radios are assumed to stay silent during the sensing interval, such that the only spectral power remaining in the air is emitted by primary users. The upper-layer protocols, e.g., the medium access control (MAC) layer, can guarantee this assumption. The duration of the sensing phase is shown by $\tau $.
\subsubsection{Local spectrum sensing}
Each CR user senses the received signal with a sampling rate $\mu $. We start from signal detection in a single narrow subcarrier, which will form a building block for our cooperative multiband detection procedure. Within duration of $\tau $ in the sensing phase, the $k^{th}$ CR user collects $\mu \tau $ samples of the received primary signal for the $\textit{n}^{th}$ subcarrier ($y_{k,n}$). Following \cite{digham2007energy}, for the $k^{th}$ CR user, its received primary signal at subcarrier $n$ in the frequency domain, $y_{k,n}$, can be given based on the following binary hypothesis:
\[{{\mathcal H}}_{1,n}:\ y_{k,n}\left(i\right)=h_{k,n}s_n\left(i\right)+v_{k,n}\left(i\right)\] 
\begin{equation}
{{\mathcal H}}_{0,n}:\ y_{k,n}\left(i\right)=v_{k,n}\left(i\right),\ \ \ 1\le i\le \mu \tau
\end{equation}
where ${{\mathcal H}}_{0,n}$ represents the absence of primary signals, and the alternative hypothesis ${{\mathcal H}}_{1,n}$ represents the presence of primary signals at subcarrier $n$; $v_{k,n}\left(i\right)$ and $h_{k,n}$ are the additive white Gaussian noise, and the complex gain of the sensing channel between the PU transmitter and the $k^{th}$ CR receiver, respectively; and $s_n\left(i\right)$ represents the signal of the primary transmitter. 

The energy detection is performed by measuring the energy of the received signal $y_{k,n}$. The energy collected in the frequency domain is denoted by $E_{k,n}$. It then follows in \cite{digham2007energy} that under ${{\mathcal H}}_{1,n}$, $y_{k,n}$ has a noncentral chi-squared distribution with variance $\sigma^2=1$, noncentrality parameter $\mu=2 \gamma$, and $2 \mu \tau$ degrees of freedom, And given ${{\mathcal H}}_{0,n}$, $y_{k,n}$ has central chi-squared distribution. Note that the instantaneous SNR of the received signal at the $k^{th}$ CR is $\gamma$. By comparing the energy $E_{k,n}$ with a threshold $\zeta_k$, the local detection of PU signal is made.

The average probability of false alarm and detection over Rayleigh fading channels for subcarrier $n$ at the $k^{th}$ CR are given by \cite{digham2007energy}, respectively

\begin{equation}
P^{(k)}_{fa}=\frac{\Gamma \left(\mu \tau, \zeta_k/2\right)}{\Gamma(\mu \tau)}
\end{equation} 
\begin{multline}
P^{(k)}_{d}=e^{-\frac{\zeta_k}{2}}\sum^{\mu \tau-2}_{p=0}{\frac{1}{p!}\left(\frac{\zeta_k}{2}\right)^p}+\left(\frac{1+\bar{\gamma_k}}{\bar{\gamma_k}}\right)^{\mu \tau-1} \\ \times \left[e^{-\frac{\zeta_k}{2(1-\bar{\gamma_k})}}-e^{-\frac{\zeta_k}{2}}\sum^{\mu \tau-2}_{p=0}{\frac{1}{p!}\left(\frac{\zeta_k \bar{\gamma_k}}{2(1+\bar{\gamma_k})}\right)^p}\right]
\end{multline} 
where $\bar{\gamma_k}$ denotes the average SNR at the $k^{th}$ CR, $\Gamma(a,x)$ is
the incomplete gamma function and $\Gamma(a)$ is the gamma function\cite{digham2007energy}.

\subsubsection{Cooperative spectrum sensing}
In cooperative spectrum sensing based on decision fusion, all CRs identify the available subcarriers of the spectrum independently, and make a binary decision on whether the $n^{th}$ subcarrier is occupied by PU or not \cite{letaief2009cooperative}.
The coordinator collects binary decisions $D_{k,n}\in \left\{0, 1\right\}$, which denotes the local spectrum sensing result of the $k^{th}$ CR at $n^{th}$ subcarrier, from all CR users.The coordinator fuses all 1-bit decisions together and makes decision on whether subcarrier $n$ is occupied by primary users or not.

It has been shown in \cite{letaief2009cooperative} that the \textit{OR} rule, in which the coordinator concludes the presence of the PU signal at the $n^{th}$ subcarrier, when there exists at least one CR that has the local decision ${{\mathcal H}}_{1,n}$, is the best among the other fusion rules. Therefore, we shall consider the \textit{OR} rule in the sequel. 

The false alarm probability and the detection probability of cooperative spectrum sensing based on decision fusion with the \textit{OR} rule for subcarrier $n$ are given respectively by \cite{letaief2009cooperative}
\begin{equation}
Q_{fa}=1-\prod^{K}_{k=1}{\left(1-P^{(k)}_{fa}\right)}
\end{equation} 
\begin{equation}
Q_{d}=1-\prod^{K}_{k=1}  \left(1-P^{(k)}_{d}\right)
\end{equation} 
It is shown in Fig. 5 of \cite{letaief2009cooperative} that for a given probability of false alarm, the missed detection probability greatly decreases when the number of cooperative CRs increase, and it refers to $K$ as the \textit{sensing diversity order} of cooperative spectrum sensing.

\subsection{Proposed Resource Allocation Framework}
Despite other recent resource allocation schemes in cognitive radio systems, in which each CR user is assigned the whole or a portion of one subcarrier at a certain transmission power, employing fragmented-spectrum synchronous OFDM-CDMA modulation exploits frequency diversity by spreading the power of each CR user on all free subcarriers. In particular, after each sensing period, the coordinator chooses an orthogonal family of codes with length equals to the total number of free subcarriers, and broadcasts the modified version of these codes, in which the corresponding chips of busy subcarriers are set to zero, to CR users. Therefore, in transmission phase ($\tau \le t\le T$), CR users spread their power using their specified codes on free subcarriers, and transmit without interfering with primary users. 
The coordinator decides to set the $n^{th}$ chip of the spreading codes to zero when one of the following two scenarios occurs:

\begin{enumerate}
\item  Subcarrier $n$ is estimated to be busy and it is indeed occupied by primary users. The probability of this scenario is ${\Pr  \left({{\mathcal H}}^1_n\right)\ }P_d$.

\item  Subcarrier $n$ is wrongly estimated to be occupied, i.e., it is actually free. This scenario happens with probability ${\Pr  \left({{\mathcal H}}^0_n\right)\ }P_{fa}$.
\end{enumerate}

Therefore, the $n^{th}$ chip of the spreading codes is set to zero with the probability
\begin{equation}
P^0_n={\Pr  \left({{\mathcal H}}^1_n\right)\ }P_d+{Pr \left({{\mathcal H}}^0_n\right)\ }P_{fa}\ \ \ \
\end{equation} 

In the rest of this paper, we assume that all subcarriers may be occupied by primary users with the same probability, i.e., $Pr\left({{\mathcal H}}^1_n\right)=Pr\left({{\mathcal H}}^1\right)$ for $1\le n\le N$.
\subsection{Proposed Real-Valued Multi-Level Orthogonal Spreading Codes}
For any positive integer $n$, we intend to find a $n\times n$ matrix $C_n$ with integer nonzero elements, such that  $C_n\times{\left(C_n\right)}^T$ becomes a diagonal matrix, i.e., the rows of $C_n$ make a set of orthogonal vectors. 
\begin{figure*}
\centering
\includegraphics[width=7in]{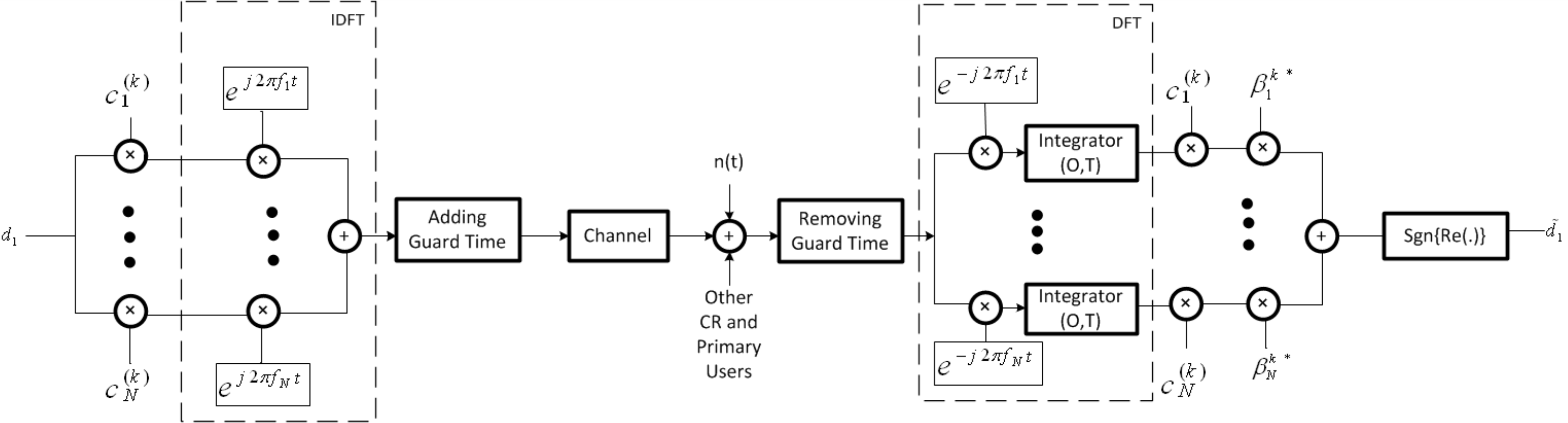}
\caption{$k^{th}$ CR user's transmitter-receiver block diagram in a typical fragmented-spectrum synchronous OFDM-CDMA system.}
\end{figure*}
\noindent If $n=2^k$, we can use Walsh algorithm, in fact $C_{2k}\ $could be produced by $C_k$ as $C_{2k}=\left( \begin{array}{cc}
C_k & C_k \\ 
C_k & {-C}_k \end{array}
\right)$. But what if $n\ \ne 2^k$ ? Here we show a novel algorithm that for any composite number $n$ ,we can produce orthogonal matrix $C_n$ . In this case, $n$ has at least one divisor other than $1$ and  $n$. Name this divisor$\ k$. Hence, $n$ could be written as $n=k.r$ such that $\ 1<k,r<n$ . Now we claim that $C_n$ could be produced by $C_k$ and $C_r$ . Define$\ \ {\ C}_n=$ $\left( \begin{array}{ccc}
a_{11}*C_k & \cdots  & a_{1r}*C_k \\ 
\vdots  & \ddots  & \vdots  \\ 
a_{r1}*C_k & \cdots  & a_{rr}*C_k \end{array}
\right)$ , in which, $\left( \begin{array}{ccc}
a_{11} & \cdots  & a_{1r} \\ 
\vdots  & \ddots  & \vdots  \\ 
a_{r1} & \cdots  & a_{rr} \end{array}
\right)=\ C_r$ .

\noindent Proof :  Here we demonstrate that $C_n\times{\left(C_n\right)}^T$ is a diagonal matrix.

\[C_n\times{\left(C_n\right)}^T=\ \left( \begin{array}{ccc}
b_{11}*C_kC^T_k & \cdots  & b_{1r}*C_kC^T_k \\ 
\vdots  & \ddots  & \vdots  \\ 
b_{r1}*C_kC^T_k & \cdots  & b_{rr}*C_kC^T_k \end{array}
\right)\] 
\[b_{ij}=\sum^r_{s=1}{a_{is}a_{js}}=\ {(C_rC^T_r)}_{ij}
\to \ \left( \begin{array}{ccc}
b_{11} & \cdots  & b_{1r} \\ 
\vdots  & \ddots  & \vdots  \\ 
b_{r1} & \cdots  & b_{rr} \end{array}
\right)=\ C_rC^T_r\] 
Since $C_kC^T_k$ and $C_rC^T_r$ are diagonal matrices, $C_nC^T_n$ is a diagonal matrix $\blacksquare $

\noindent Therefore, if we find $C_n$ for all prime numbers, we are able to find $C_n$ for all numbers.

\noindent Also note that if $\ r=2$ , $C_r=\ \left( \begin{array}{cc}
1 & 1 \\ 
1 & -1 \end{array}
\right)$ ,therefore $C_n=\ \left( \begin{array}{cc}
C_k & C_k \\ 
C_k & {-C}_k \end{array}
\right)$ and this is the Walsh Codes of length $n=2k$ .
\subsection{ Cognitive Radio User's Signal Model}
Fig. 1 shows $k^{th}$ CR user's transmitter-receiver block diagram in a typical fragmented-spectrum synchronous OFDM-CDMA system. As shown in the figure, the modified signature code $\{c^{\left(k\right)}_n,\ n=1,2,\ \dots ,N\}$ of the $k^{th}$ user is spread among a set of orthogonal subcarriers, each carrying the same information bit. Thus, each chip modulates one of the orthogonal subcarriers. In other words, $c^{\left(k\right)}_n$ takes \{$\pm $1, $\pm $2, $\pm $3, $\pm $4. ...\} for free subcarriers and \{0\} for busy ones. This process is applied by the $N$-point IDFT block. A guard time is then added to reduce the effects of ISI. The subcarriers are separated by $1/T_b$ and the two-sided null-to-null bandwidth associated with each subcarrier is $\Omega =2/T_b$ where it should be selected to be smaller than the coherence bandwidth $B_c$, so that each subcarrier encounters flat fading. For simplicity, we will consider our reference time to be zero instead of $\tau $ in the transmission phase. Thus, the transmitted signal of the $k^{th}$ CR user for transmitting $I$ bits in can be written as;

\begin{equation}
s^{\left(k\right)}_{Tx}\left(t\right)=\sum^I_{i=1}{\sum^N_{n=1}{d^k_ic^{\left(k\right)}_n\sqrt{P_n}e^{j2\pi f_nt}q\left(t-\left(i-1\right)T_b\right)}}
\end{equation}
where  $q\left(t\right)=1/\sqrt{T_b},\ \ 0<t<T_b$ is the normalized symbol waveform, $d^k_i\in \{\pm 1\}$ is the $i^{th}$ transmitted bit, and $f_n=n/T_b$ is the $n^{th}$ subcarrier frequency. As we mentioned previously, in the fragmented-spectrum synchronous OFDM-CDMA technique, transmitted power of each CR user is spread on all free subcarriers. If $\varepsilon_b$ represents transmitted power per bit, each CR user devotes $P_n=\frac{\varepsilon_b}{\sum^N_{n=1}\left(c^{(k)}_n\right)^2}$ to free subcarriers and $P_n=0$ to busy ones. Note that since $\sum^N_{n=1}\left(c^{(k)}_n\right)^2$ is the same for all $k=1, ..., K$, $P_n$ is independent of $k$. At the receiver, the guard time is discarded and the signal is fed to an $N$-point DFT block to separate the subcarriers. The received signal for the first symbol can be written as
\begin{multline}
r\left(t\right)=\sum^K_{k=1}{\sum^N_{n=1}{{\beta }^k_nd^kc^{(k)}_n\sqrt{P_n}e^{j2\pi f_nt}q\left(t\right)}} \\+ \sum_{j\in \Lambda }{s_j\left(t\right)e^{j2\pi f_jt}}+n\left(t\right),\ \ \ \ \ \ \ \ 0<t<T_b
\end{multline}

\noindent in which ${\beta }^k_n$ is the channel gain of the $n^{th}$ subcarrier between the $k^{th}$ CR transmitter and its receiver, and is assumed to be \textit{i.i.d.} circularly symmetric complex Gaussian (CSCG) random variable with zero mean and unit variance; $n\left(t\right)$ is also additive white CSCG noise with power spectral density ${\sigma }^2_n$ \cite{liang2008sensing}. We define a set $\Lambda $ which contains $l$ indices that denote those subcarriers in which misdetection has been occurred. In other words, these subcarriers are occupied by primary users but we wrongly estimate them to be free. $s_j\left(t\right)$ denotes primary signal interference to the CR receiver on the $j^{th}$ subcarrier and can be modeled as a Gaussian random process with variance ${\sigma }^2_s$ \cite{quan2009optimal}.

\section {Cognitive Radio Receiver Performance Analysis}

In this section, we analyze the performance of the first CR receiver based on the proposed fragmented-spectrum synchronous OFDM-CDMA Technique. Define the test statistics to be the projection of the received signal on the Fourier basis functions. The $n^{th}$ test statistic is
\begin{multline}
r_n=\frac{1}{\sqrt{T_b}}\int^{T_{b\ }}_0{r\left(t\right)}e^{-j2\pi f_nt}dt=\sqrt{P_n}d^1c^{\left(1\right)}_n{\beta }^{(1)}_n\\+\sqrt{P_n}\sum^K_{k=2}{{\beta }^k_nd^kc^{\left(k\right)}_n}+u_ns_n+v_n
\end{multline}
where ${\{v}_n\}$ and $\{s_n\}$ are uncorrelated Gaussian random variables with zero mean and variance ${\sigma }^2_n$ and ${\sigma }^2_s$, respectively; and $u_n=1$ for  $n\in \Lambda $ and $u_n=0$ for others. 

\noindent With perfect channel estimation, the decision variable can be written as
\begin{equation}
R=\sum^N_{n=1}{\sqrt{P_n}Re\{{{\beta }^1_n}^*c^{\left(1\right)}_nr_n\}}=R_s+R_{MAI}+R_{GI}+R_n
\end{equation}
where $*$ denotes complex conjugate operation; $R_s, R_{MAI}, R_{GI}$ and $R_n$ indicate the desired information bit term, multiple-access interference term (MAI), Gaussian interference term (GI) due to primary user misdetection, and additive white Gaussian noise term (AWGN), respectively. The desired information bit term is evaluated as
\begin{equation}
R_s=d^1{\sum^N_{n=1}{{\left|{\beta }^{(1)}_n\right|}^2}{\left(c^{\left(1\right)}_n\right)^2}P_n} 
\end{equation}
According to the central limit theorem, for large values of $N$, we can approximate $R_s$ by a Gaussian random variable. Its mean and variance can be obtained (by substituting $P_n=\frac{\varepsilon_b}{\sum^N_{n=1}\left(c^{(1)}_n\right)^2}$ for free subcarriers and $P_n=0$ for others) as $d^1{\varepsilon }_b$ and  ${{\left(d^1\right)^2}}{{\varepsilon }_b}^2\frac{\sum^N_{n=1}\left(c^{(1)}_n\right)^4}{\left({\sum^N_{n=1}\left(c^{(1)}_n\right)^2}\right)^2}$, respectively. The multiple-access interference term in decision variable can be derived as
\begin{multline}
R_{MAI}=\\ \sum^N_{n=1}{P_n\sum^K_{k=2}{\left(Re\{{\beta }^1_n\}{Re\{\beta }^k_n\}+{Im\{\beta }^1_n\}{Im\{\beta }^k_n\}\right)d^kc^{\left(1\right)}_nc^{\left(k\right)}_n}}
\end{multline}
where $Re\{{\beta }^1_n\}$ and $Im\{{\beta }^1_n\}$ denote real and imaginary parts of  ${\beta }^1_n$, respectively. According to the central limit theorem, $R_{MAI}$ can be modeled as
\begin{equation}
R_{MAI}\sim N(0,\frac{1}{2}{{\varepsilon }_b}^2\frac{\sum^K_{k=2}{\sum^N_{n=1}\left({c^{(1)}_n}{c^{(k)}_n}\right)^2}}{{\left({\sum^N_{n=1}\left(c^{(1)}_n\right)^2}\right)^2}})
\end{equation}
The effect of the primary user signal in the desired CR receiver due to misdetection is obtained as
\begin{equation}
R_{GI}=\sum_{n\in \Lambda }{\sqrt{P_n}c^{\left(1\right)}_n{(Re\{\beta }^1_n\}Re\left\{s_n\right\}+{Im\{\beta }^1_n\}Im\{s_n\})}
\end{equation} 
We can also approximate $R_{GI}$ with a Normal random variable as
\begin{equation}
R_{GI}\sim N(0,\frac{1}{2}{\frac{{\varepsilon }_b}{\sum^N_{n=1}\left(c^{(1)}_n\right)^2}}{\left(\sum_{n\in \Lambda }{\left(c^{(1)}_n\right)^2}\right)}\sigma_s^2)
\end{equation} 
Finally, the AWGN term in the decision variable is equal to
\begin{equation}
R_n=\sum^N_{n=1}{\sqrt{P_n}c^{\left(1\right)}_n{\ (Re\{\beta }^1_n\}Re\left\{v_n\right\}+{Im\{\beta }^1_n\}Im\{v_n\})}
\end{equation}
which can be similarly modeled as $R_n\sim N(0,\frac{1}{2}{\varepsilon }_b{\sigma }^2_n)$ for large values of N.
\begin{figure}
\centering
\includegraphics[width=3.5in]{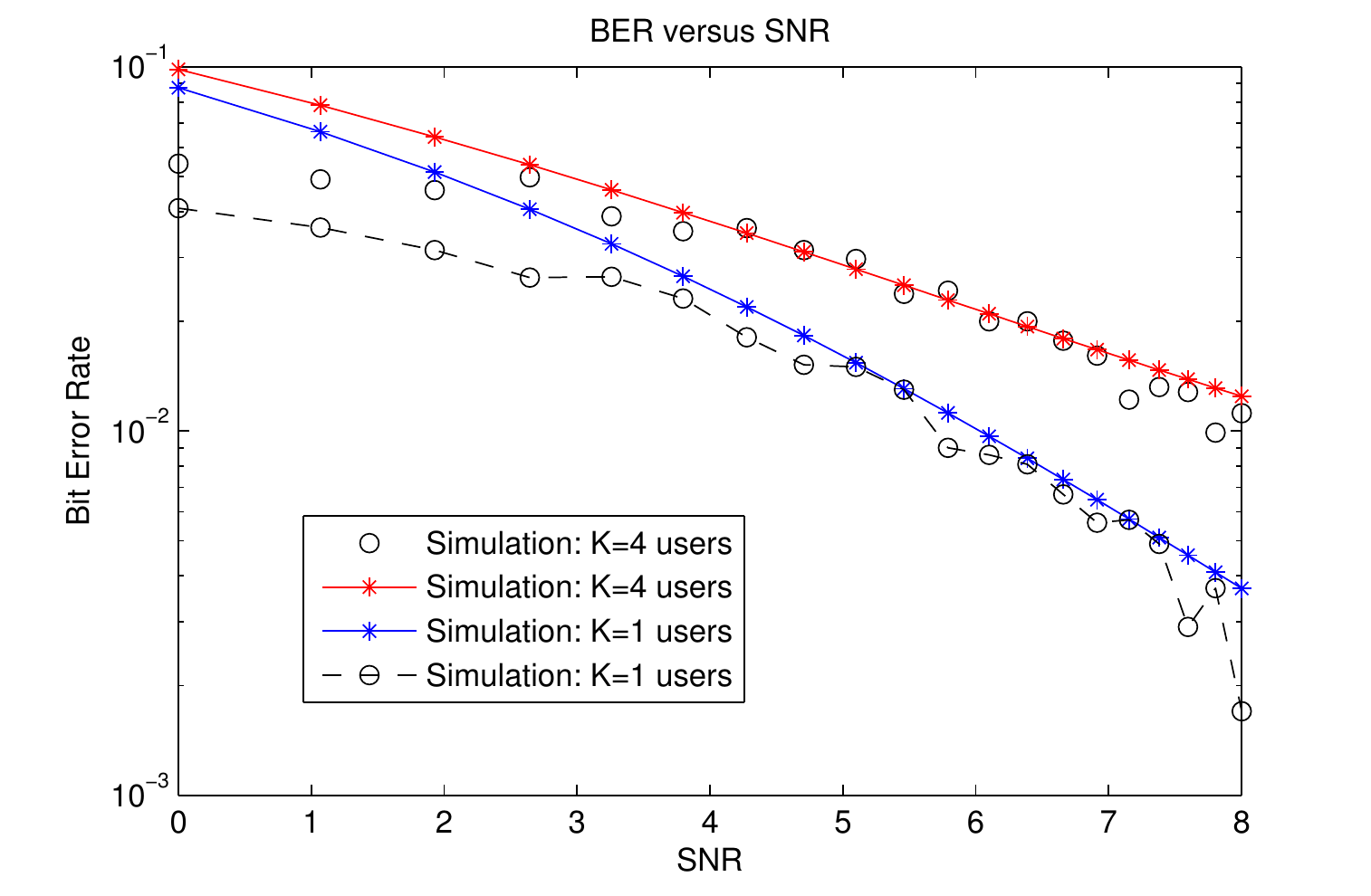}
\caption{		Bit Error Rate versus CR users' $SNR$ for two values of $ K $. $ N=32,  P_d=0.95., Pr\left({{\mathcal H}}^1\right)=0.2$.}
\end{figure}

\begin{figure}
\centering
\includegraphics[width=3.5in]{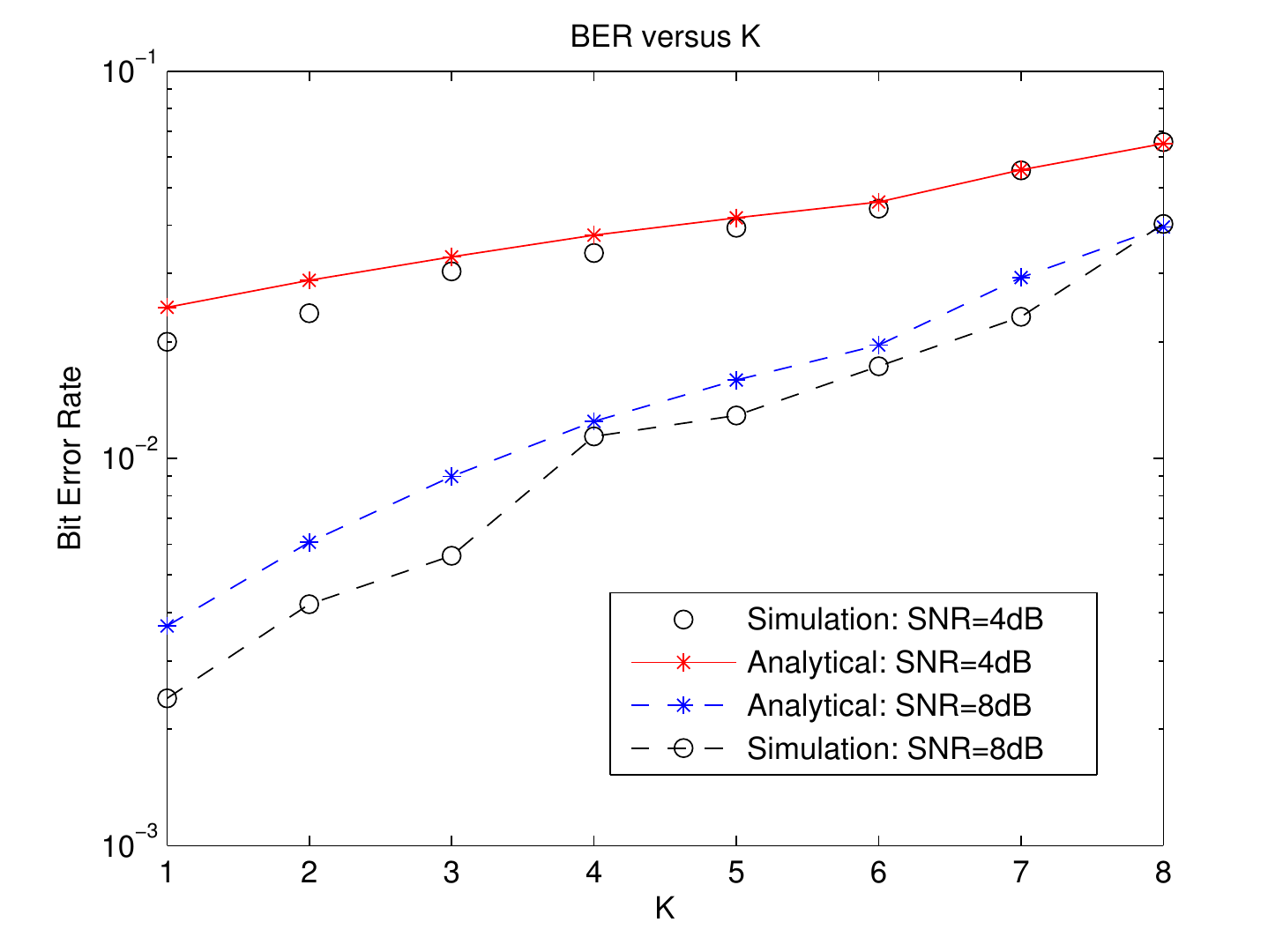}
\caption{	Bit Error Rate versus number of CR users ($K$) for two values of $SNR$. $N=32, P_d=0.95, Pr\left({{\mathcal H}}^1\right)=0.2$.}
\end{figure}
\noindent Without any loss of generality we assume the first user to transmit data bit ``1''. Thus if we consider $m$ as the number of subcarriers which are estimated to be occupied, and let $l$ be the number of subcarriers which are occupied by primary users but the coordinator have failed to detect them, the conditional probability of error can be derived as
\begin{multline}
P_{e|l,m}=P\left(R<0\right)= \\ Q( \frac{{\varepsilon }_b}{\sqrt{var_s+var_{MAI} +var_{GI}+var_n}})
\end{multline}
The values of $var_s$, $var_{MAI}$, $var_{GI}$,and $var_n$ are as follows:
\begin{equation}
var_s = {{\varepsilon }_b}^2\frac{\sum^N_{n=1}\left(c^{(1)}_n\right)^4}{\left({\sum^N_{n=1}\left(c^{(1)}_n\right)^2}\right)^2}
\end{equation}
\begin{equation}
var_{MAI} = \frac{1}{2}{{\varepsilon }_b}^2\frac{\sum^K_{k=2}{\sum^N_{n=1}\left({c^{(1)}_n}{c^{(k)}_n}\right)^2}}{{\left({\sum^N_{n=1}\left(c^{(1)}_n\right)^2}\right)^2}}
\end{equation}
\begin{equation}
var_{GI} = \frac{1}{2}{\frac{{\varepsilon }_b}{\sum^N_{n=1}\left(c^{(1)}_n\right)^2}}{\left(\sum_{n\in \Lambda }{\left(c^{(1)}_n\right)^2}\right)}\sigma_s^2
\end{equation}
\begin{equation}
var_n = \frac{1}{2}{\varepsilon }_b\sigma^2_n
\end{equation}
The probability that the $n^{th}$ subcarrier is decided to be busy is $P^0_n$, which is obtained in (6), and the probability that misdetection occurs in $n^{th}$ subcarrier is equal to $P_{mis}=\left(1-P_d\right)Pr\left({{\mathcal H}}^1\right)$.

\noindent By averaging on $m$ and $l$, the probability of error can be written as
\begin{multline}
P_e=\sum^N_{m=1}{\sum^{N-m}_{l=1}{{\left( \begin{array}{c}
N \\ 
m \end{array}
\right)\left( \begin{array}{c}
N-m \\ 
l \end{array}
\right)\left\{P^0_n\right\}^m}}} \\ \times {\left\{P_{mis}\right\}^l}{\left\{1- P^0_n -P_{mis} \right\}^{N-m-l}} \\ \times Q( \frac{{\varepsilon }_b}{\sqrt{var_s+var_{MAI} +var_{GI}+var_n}})
\end{multline}

\section{ Numerical Results}

In this section, we numerically evaluate the proposed fragmented-spectrum synchronous OFDM-CDMA framework. Consider a primary system with $3.2$ MHz bandwidth which is equally divided into $N=32$ subbands. The CR user bit duration time is $T_b=\frac{N}{BW}=10\ \mu s$; the sampling rate is $\mu =3.2$ MHz and the slot duration is $T=1\ ms$. All links (from primary to CR users and between CR users) experience Rayleigh fading. Furthermore, the channel $SNR$ value from the primary user to a CR user has a mean of 2.3 dB. The threshold value, $\varepsilon $, is set such that in all scenarios, $P_d$ achieves the desired value.


We set the value of $\tau $ large enough so that false alarm probability is negligible for the desired detection probability ($P_d=0.95$). Fig. 2 and 3 represent the probability of error in terms of CR users' signal to noise power ratio and various $K$. It can be inferred from the figures that we can have a relatively reliable fragmented-spectrum synchronous OFDM-CDMA system for our cognitive radio network with 4 users. While, by increasing the number of users to 8 , an increase in $SNR$ will not lead to considerable performance improvement. The simulation curves in these figures lay approximately on their corresponding analytical curves which validate our analysis.

%


\section{Conclusion}
\noindent In this paper, a novel resource allocation method based on fragmented-spectrum synchronous OFDM-CDMA modulation has been proposed to simplify multiple access capability in cognitive radio (CR) systems. This framework also improves the performance by exploiting frequency diversity. In particular, a spectrum sensing method was presented to detect preexisting primary users (PU) in a target bandwidth and an algorithm was used to assign CR users OFDM-CDMA signature codes in order to prevent interfering with PUs. Furthermore, closed form expression for bit error probability of the cognitive radio receivers was derived. Based on numerical results, it was shown that the proposed scheme in this paper offers an excellent resource allocation for cognitive radio networks.








%
\bibliographystyle{IEEEtranTCOM}
\bibliography{IEEEabrv}

\end{document}